\documentclass[aps,prl,twocolumn,superscriptaddress]{revtex4}

\usepackage[a4paper, total={7in, 9in}]{geometry}

\usepackage{graphicx}
\usepackage{amsmath}
\usepackage{soul}
\usepackage{color}
\usepackage{textcomp}
\usepackage{booktabs}
\usepackage[flushleft]{threeparttable}
\usepackage{siunitx}
\usepackage{amssymb}
\usepackage{amsmath}
\usepackage{float}
\usepackage{psfrag}
\usepackage{siunitx}
\usepackage{ulem}
\usepackage{romannum}
\usepackage{comment}

\def\be{\begin{equation}}
\def\ee{\end{equation}}
\def\ba{\begin{eqnarray}}
\def\ea{\end{eqnarray}}

\begin{document}

\title{A bouncing oil droplet in a stratified liquid and its sudden death}

\author{Yanshen Li}
\affiliation{Physics of Fluids group, Max-Planck Center Twente for Complex Fluid Dynamics, Department of Science and Technology, Mesa+ Institute, and 
J. M. Burgers Centre for Fluid Dynamics, University of Twente, P.O. Box 217, 7500 AE Enschede, The Netherlands}
\author{Christian Diddens}
\affiliation{Physics of Fluids group, Max-Planck Center Twente for Complex Fluid Dynamics, Department of Science and Technology, Mesa+ Institute, and 
J. M. Burgers Centre for Fluid Dynamics, University of Twente, P.O. Box 217, 7500 AE Enschede, The Netherlands}
\affiliation{Department of Mechanical Engineering, Eindhoven University of Technology, P.O. Box 513, 5600 MB Eindhoven, The Netherlands}
\author{Andrea Prosperetti}
\affiliation{Physics of Fluids group, Max-Planck Center Twente for Complex Fluid Dynamics, Department of Science and Technology, Mesa+ Institute, and 
J. M. Burgers Centre for Fluid Dynamics, University of Twente, P.O. Box 217, 7500 AE Enschede, The Netherlands}
\affiliation{Department of Mechanical Engineering, University of Houston, TX 77204-4006, USA}
\author{Kai Leong Chong}
\affiliation{Physics of Fluids group, Max-Planck Center Twente for Complex Fluid Dynamics, Department of Science and Technology, Mesa+ Institute, and 
J. M. Burgers Centre for Fluid Dynamics, University of Twente, P.O. Box 217, 7500 AE Enschede, The Netherlands}
\author{Xuehua Zhang}
\email{xuehua.zhang@ualberta.ca}
\affiliation{Department of Chemical and Materials Engineering, University of Alberta, 12-211 Donadeo
Innovation Centre for Engineering, Edmonton, Alberta, Canada}
\affiliation{Physics of Fluids group, Max-Planck Center Twente for Complex Fluid Dynamics, Department of Science and Technology, Mesa+ Institute, and 
J. M. Burgers Centre for Fluid Dynamics, University of Twente, P.O. Box 217, 7500 AE Enschede, The Netherlands}
\author{Detlef Lohse}
\email{d.lohse@utwente.nl}
\affiliation{Physics of Fluids group, Max-Planck Center Twente for Complex Fluid Dynamics, Department of Science and Technology, Mesa+ Institute, and 
J. M. Burgers Centre for Fluid Dynamics, University of Twente, P.O. Box 217, 7500 AE Enschede, The Netherlands}
\affiliation{Max Planck Institute for Dynamics and Self-Organization, 37077 G\"ottingen, Germany}

\begin{abstract}
Droplets can self-propel when immersed in another liquid in which a concentration gradient is present. Here we report the experimental and numerical study of a self-propelling oil droplet  in a vertically stratified ethanol/water mixture: At first, the droplet sinks slowly due to gravity, but then, before having reached its density matched position, jumps up suddenly. More remarkably, the droplet bounces repeatedly with an ever increasing jumping distance, until all of a sudden it stops after about \SI{30}{\minute}. We identify the Marangoni stress at the droplet/liquid interface as responsible for the jumping: its strength grows exponentially because it pulls down ethanol-rich liquid, which in turn increases its strength even more. The jumping process can repeat because gravity restores the system. Finally, the sudden death of the jumping droplet is also explained. Our findings have demonstrated a type of prominent droplet bouncing inside a continuous medium with no wall or sharp interface.
\end{abstract}


\maketitle
Swimming droplets \cite{maass2016swimming} are of great importance for their relevance to (bio)chemical reactors \cite{demello2006control, baraban2011millifluidic}. They also serve as a model system for studying collective behavior in biological populations \cite{toner1998flocks, wensink2012emergent, wensink2012meso, buttinoni2013dynamical, bialke2013microscopic}. One of the fundamental mechanisms leading to their self-propulsion is the so-called Marangoni effect \cite{scriven1960scriven, yang2018diffusiophoresis}. It is induced by the non-uniform interfacial tension of the droplet which can be generated by chemical reactions \cite{schmitt2013swimming,michelin2013spontaneous,yoshinaga2012drift,hanczyc2007fatty,toyota2009self}, solubilization \cite{kovalchuk2006marangoni, nagai2005mode, chen2009self, pena2006solubilization, izri2014self, maass2016swimming}, phase separation \cite{tanaka1998spontaneous, vladimirova1999diffusiophoresis, poesio2009dissolution, tan2016evaporation, li2018evaporation}, or by a global temperature/solute gradient \cite{kim2017solutal}. The present work focuses on the last type which is commonly encountered in nature \cite{sigman2004polar}.

    A major focus of earlier studies is on the dynamics of swimming droplets \cite{izri2014self}. The motion of a droplet in a global solute gradient is believed to be governed by the competition of droplet speed and the diffusivity of the background concentration field, which is characterized by the P\'eclet number Pe, which is the ratio between the diffusive and the inertial time scale. Previous works identified two regimes of droplet motion based on the framework of diffusiophoresis \cite{young1959motion, anderson1989colloid, yang2018diffusiophoresis}: For small Pe, the concentration gradient is not affected by fluid motion and thus the droplet movement can persist. In contrast, for large Pe, the sharp concentration gradient at the periphery of the droplet is always smoothed out, and thus the motion of the droplet is slowed down \cite{michelin2014phoretic}. However, in this Letter, we conduct experiments and simulations to demonstrate that when combined with gravity, the Marangoni stress on an oil droplet in a stratified ethanol/water mixture can oscillate between large and small Pe, leading to a continuous bouncing of the droplet. And, more surprisingly, the amplitude of the droplet oscillatory motion even increases before it suddenly stops. Contrary to the commonly held concept that droplet bouncing requires a wall \cite{richard2002surface, bouwhuis2012maximal, shirota2016dynamic} or a sharp interface \cite{blanchette2012drops}, here the droplet bounces in the bulk of a continuous medium, which only requires a large enough concentration gradient. 

\begin{figure*}[t!] 
	\centering
	\setlength{\unitlength}{\textwidth}
	\includegraphics[width=0.9\unitlength]{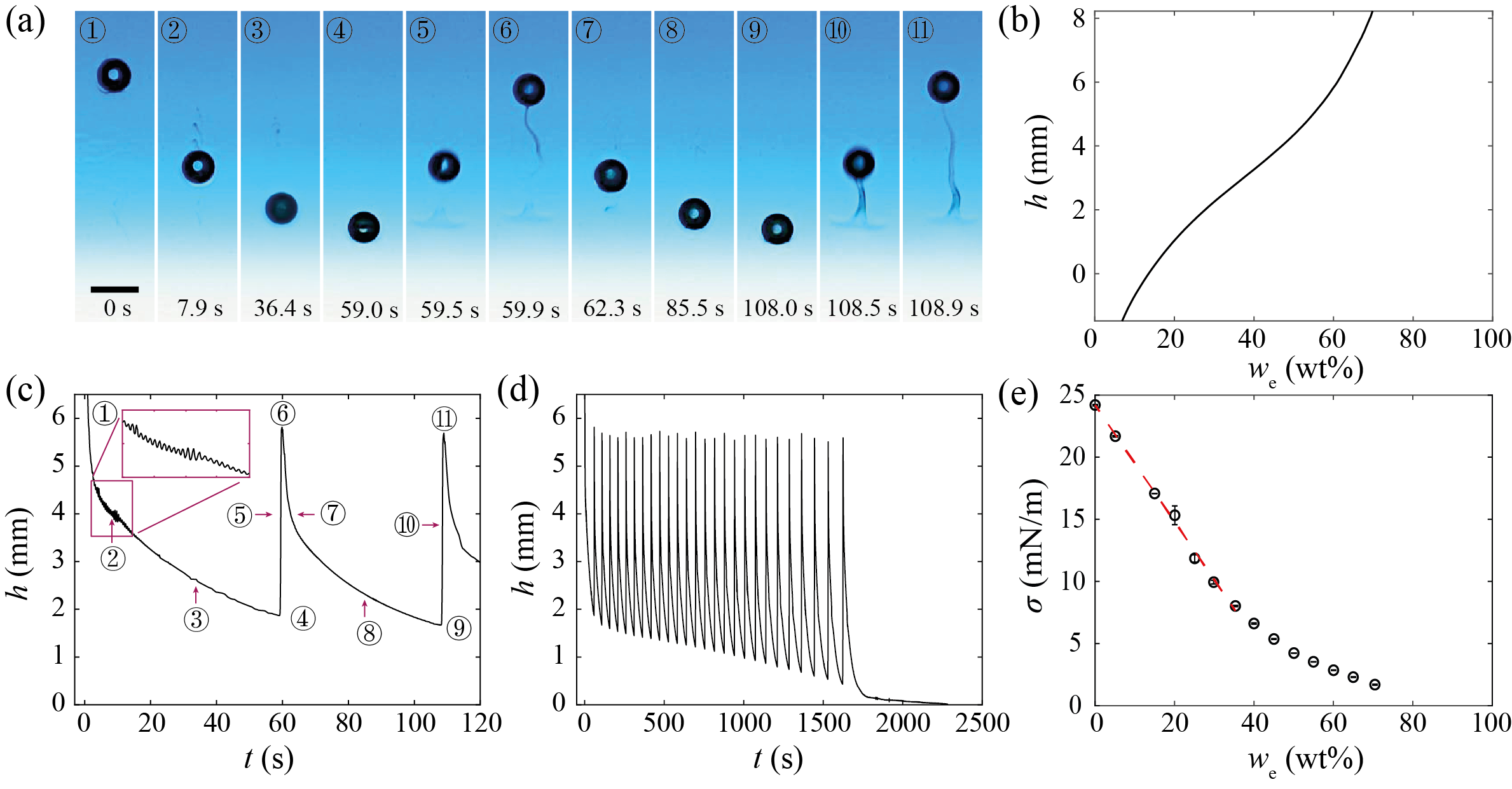}
\caption{Continuous jumps and the eventual sudden death of the oil drop. (a) Successive snapshots of the \SI{0.5}{\micro\liter} oil drop for the first two cycles. Ethanol is dyed blue. The scale bar is \SI{1}{mm}. (b) Ethanol weight fraction $w_\text{e}$ as a function of height $h$ is measured by laser deflection. (c) The oil drop's centre position $h$ versus time $t$ for the first 2 cycles, with the final height being {taken as} 0.  The insert shows the vertical oscillation, with frequency \SI{2.15}{\hertz}, {during the sinking of the drop}. (d) The oil drop's centre position $h$ for all the cycles of the jumping process. After each jump, the drop sinks to a lower position, but at each jump, it still reaches almost the same height $h\approx \SI{6}{mm}$, thus even increasing the jumping amplitude. This particular drop jumps 26 times within \SI{30}{\minute}. (e) The interfacial tension between oil and ethanol/water mixture. Error bars are the standard deviation over 5 measurements.} 
\label{fig:1}
\end{figure*}

In the experiment, \SI{1.8}{mL} ethanol is carefully injected into a cuvette ($10\times10\times \SI{45}{mm}$) containing \SI{1.8}{mL} water to produce a vertical density stratification. Then a \SI{0.5}{\micro\liter} ($R = \SI{0.44}{mm}$) oil drop of trans-Anethole is released in it. The motion of the droplet is visualized with a Nikon camera aiming from the side. A series of typical snapshots of the droplet motion within the first two jumping cycles are shown in Fig.\ref{fig:1}(a). Ethanol is dyed blue (Methylene Blue Hydrate) to visualize the concentration gradient, and the ethanol fraction $w_\mathrm{e}$ as a function of height $h$ is measured by laser deflection and shown in Fig.\ref{fig:1}(b). The height of the droplet center $h(t)$ is plotted over the first two cycles in Fig.\ref{fig:1}(c) and over the entire jumping lifetime in Fig.\ref{fig:1}(d), with its final value being taken as 0. 

The oil droplet has a density of \SI{988}{kg/m^3} at \SI{25}{\degreeCelsius}, which is slightly lighter than water (\SI{997}{kg/m^3}) but much denser than ethanol (\SI{785}{kg/m^3}), so it first sinks slowly due to gravity (Fig.\ref{fig:1}(a), {\textcircled{\scriptsize 1}}-{\textcircled{\scriptsize 3}}). At \SI{59}{\second} ({\textcircled{\scriptsize 4}}), the droplet reaches the height with surrounding mixture density $\rho_\mathrm{mix}=\SI{958}{kg/m^3}$ ($w_\text{e}=\SI{26.6}{wt\%}$), which is still lighter than the oil droplet. Surprisingly, instead of sinking continuously, the drop suddenly changes direction and jumps up by $\sim \SI{4}{mm}$ ({\textcircled{\scriptsize 4}}-{\textcircled{\scriptsize 6}}), which is more than 4 times of its diameter. It reaches the highest position within \SI{0.9}{s} ({\textcircled{\scriptsize 6}}), then sinks again for another \SI{49}{s} ({\textcircled{\scriptsize 6}}-{\textcircled{\scriptsize 9}}). Before reaching the density matched position ($h = \SI{0}{mm}$, $\rho_\mathrm{mix} = \SI{978}{kg/m^3}$) again, the drop suddenly jumps up from \SI{108}{s} ({\textcircled{\scriptsize 9}}, $w_\mathrm{e}=\SI{24.8}{wt\%}$ and $\rho_\mathrm{mix}=\SI{961}{kg/m^3}$) till \SI{108.9}{s} ({\textcircled{\scriptsize 11}}). It continuously sinks and jumps for another 24 times, then all of a sudden it falls dead after \SI{30}{\minute},  
as shown in Fig.\ref{fig:1}(d). It is noteworthy that by each jump, the drop sinks to a lower position but still returns to almost the same height ($h \approx \SI{6}{mm}$, $w_\mathrm{e} \approx \SI{60}{wt\%}$), thus the jumping distance increases progressively from \SI{4}{mm} to \SI{5.5}{mm}. We also note that during the sinking of the droplet, there are tiny oscillations as shown in the insert of Fig.\ref{fig:1}(c). These oscillations display the Brunt-V\"ais\"al\"a frequency which describes the vertical oscillation of a fluid parcel in a vertically stratified fluid around its stable position \cite{torres2000flow, hanazaki2009jets, hanazaki2009schmidt, yick2009enhanced}. Indeed, the calculated Brunt-V\"ais\"al\"a frequency in our case is \SI{2.08}{\hertz}, which fits well with the observed value \SI{2.15}{\hertz}.

\begin{figure*}[t!] 
\centering
\includegraphics[width=1\textwidth]{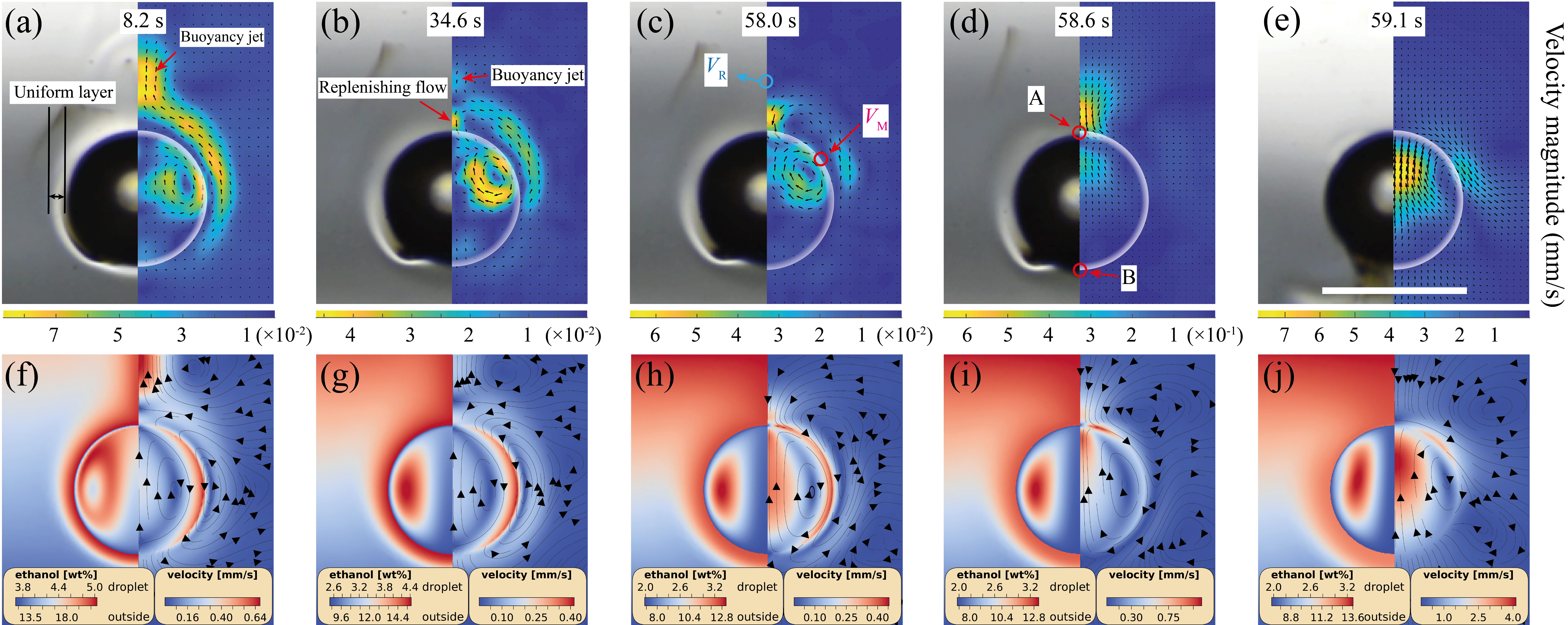}
\caption{Shadowgraph (left) and PIV (right) measurements of a \SI{0.5}{\micro\liter} oil drop (top row, (a)-(e)) and the numerical simulations (bottom row, (f)-(j)) during {the drop's} first jump. Light intensity gradient in the background of the shadowgraph {indicates} the ethanol {concentration} in the surrounding liquid. Velocities are shown in the laboratory frame. Scale bar is \SI{1}{mm}, and color bars denotes velocity magnitude as well as ethanol fraction. (a),(f) Shortly after release, {the buoyant liquid in the droplet's wake generates a relatively strong jet – the buoyancy jet.} A drifted uniform layer of ethanol-rich liquid leads to a very weak Marangoni flow.  (b),(g) {T}he Marangoni flow becomes relatively strong, thus the flow directly above the drop is pointing downwards; we refer to it as ``replenishing flow''. {Frames (c),(h) are close to the time} when the buoyancy jet vanishes. Ethanol-rich liquid above the drop is being brought downwards to its apex, and $V_\text{M}$ starts to increase. (d),(i) This downward flow brings more ethanol to the apex, further increasing $V_\textrm{M}$ in the upper half of the droplet, which then will move as a ``puller'' \cite{maass2016swimming}. (e),(j) The Marangoni flow has increased by two orders of magnitude in less than \SI{1}{\second}, pulling the drop upwards.}
\label{fig:2}
\end{figure*}

Oil has smaller surface tension with ethanol than with water (Fig.\ref{fig:1}(e)), so the ethanol-rich liquid above makes the interfacial tension at the apex of the droplet smaller than that of its bottom. This interfacial tension difference generates a Marangoni flow pointing downwards, which then {tends} to lift the droplet \cite{maass2016swimming}. The stronger the Marangoni flow, the faster the droplet will move to the opposite direction. The induced Marangoni flow is {essential} to the jumping of the droplet, and we confirm this by adding surfactant {to the bulk} liquid (\SI{0.7}{mM} Sodium dodecyl sulfate, SDS, in both water and ethanol) to suppress the Marangoni stress. {With this addition}, the droplet only sinks without jumping. {This result suggests that during the sinking motion of the droplet,} the Marangoni flow is very small.

Apart from the weak Marangoni flow on the droplet induced by the ethanol gradient, it is also settling through a density gradient. As is well known \cite{eames2003fluid, katija2009viscosity, nawroth2014induced}, a settling particle in a vertically stratified liquid brings lighter liquid down with it, the so called ``drift"  or ``entrainment''. Buoyancy of the entrained liquid acts as an extra drag on the droplet, making it sink {monotonically} towards the density matched position, i.e., no oscillation. Some of the entrained liquid will also go up, forming the ``buoyancy jet'' \cite{torres2000flow, hanazaki2009jets, hanazaki2009schmidt, yick2009enhanced}. The same concept {can} be {adapted} to the droplet in our case, except that an extra Marangoni flow, pointing downwards, is superimposed to the flow field of {the settling motion}. The resulting flow field {is} then determined by the relative strength {of} these two effects. 

We then perform PIV measurements and shadowgraphy to reveal the flow dynamics during the sinking-jumping process, as shown in the right and left panels of Fig.\ref{fig:2}(a)-(e), respectively. Shadowgraph provides qualitative information on density variations {which modulate} light intensity. In our case, the ethanol gradient in the surrounding liquid is indicated by light intensity gradient in the background, with brighter regions representing higher ethanol concentration. For the same {reason}, the drifted layer around the drop in Fig.\ref{fig:2}(a)-(d) is found to have almost uniform light intensity, meaning almost uniform ethanol concentration. The resulting Marangoni flow is very weak $V_\mathrm{M}\approx \SI{0.02}{mm/s}$ (in droplet reference frame, smaller than the sinking velocity $V_\mathrm{sink}\approx \SI{0.04}{mm/s}$), so that the droplet could sink.

Additional insight in the phenomenon can be obtained by numerical simulation. We use an axisymmetric sharp-interface finite element method with an arbitrary Eulerian-Lagrangian approach. The numerical model considers the flow and the advection and diffusion of the composition inside and outside the droplet, mass transfer by dissolution, buoyancy effects within the Boussinesq approximation and Marangoni flow. The model is implemented with the finite element package \textsc{oomph-lib}\cite{heil2006oomph}. Performed in a linear gradient, the numerical results are found to qualitatively fit the experimental results (Fig.\ref{fig:2}(f)-(j); see Supplemental Material for more details and results).  

From both experiment and numerics we conclude that the {velocity $V_\text{buo}$ of the} buoyancy jet is quite strong shortly after the drop is released (Fig.\ref{fig:2}(a),(f)), then it decreases, so that the Marangoni stress becomes relatively stronger (Fig.\ref{fig:2}(b),(g)), forming a downwards replenishing flow above the apex of the drop. This local recirculation close to the drop only decreases the Marangoni strength slowly (Pe $\sim10$, a moderate advection). Sinking deeper, the buoyancy jet becomes so weak until finally it vanishes, and the Marangoni-induced replenishing flow dominates. At this moment (Fig.\ref{fig:2}(c),(h)), {the buoyanct flow stops and the liquid velocity} above the drop is {entirely downward}. Different from the local recirculation, this downward flow brings ethanol-richer liquid to the apex of the drop, decreases the local surface tension, thus increasing the Marangoni flow (especially in the upper half), which in turn brings more ethanol, forming a positive feedback (Fig.\ref{fig:2}(d)-(e),(i)-(j)). The Marangoni flow consequently strongly increases until it is large enough to pull the droplet up. 

\begin{figure}[t!]
\centering
\includegraphics[width=0.48\textwidth]{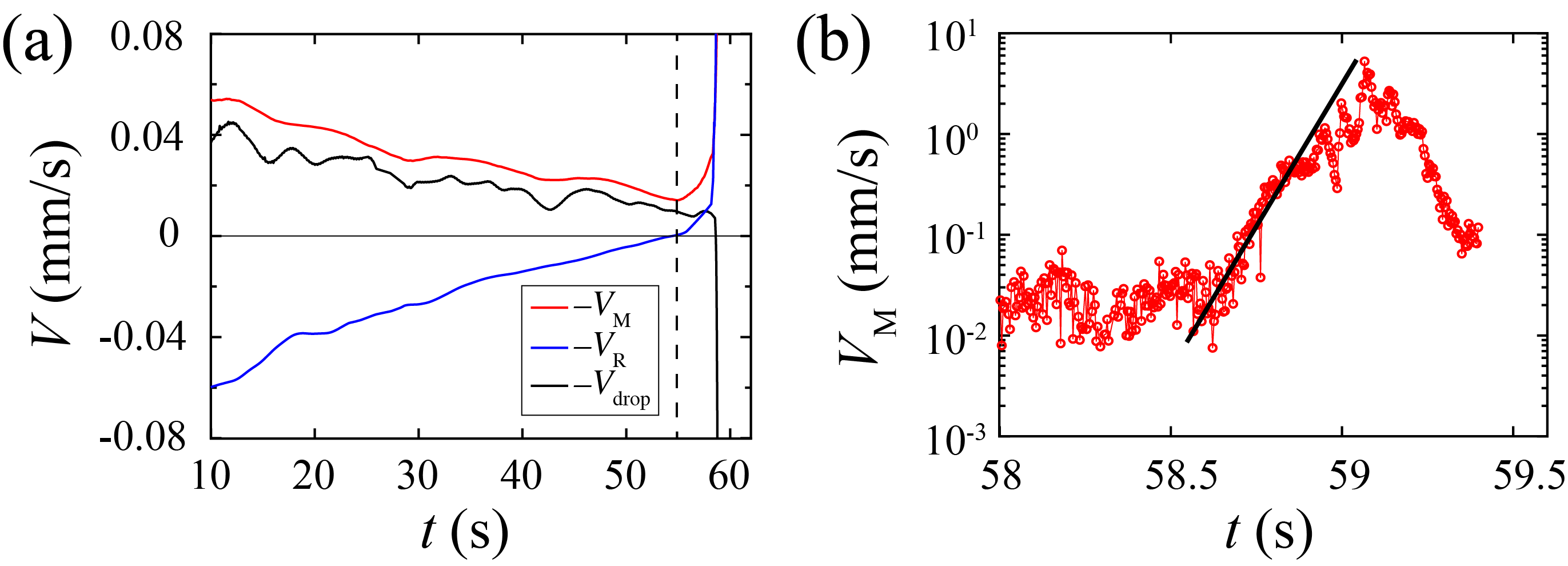}
\caption{Measured velocities as a function of time. (a) $V_\textrm{M}$ (red), $V_\textrm{R}$ (blue) and sinking velocity $V_\textrm{drop}$ (black) during the first jumping {cycle}. $V_\textrm{M}$ and $V_\textrm{R}$ are measured at the positions shown in Fig.\ref{fig:2}(c). {Positive velocity is upward}. The absolute value of $V_\textrm{M}$, $V_\textrm{drop}$ and $V_\textrm{R}$ all decrease. $V_\textrm{R}$ vanishes and then changes direction. This leads to the {sharp} increase of $V_\textrm{M}$ {around $t\approx \SI{55}{s}$}, until it is large enough to pull the droplet up.  (b) $V_\textrm{M}$ in logarithmic scale as a function of time in the later stage. $V_\textrm{M}$ increases exponentially between \SI{58.6}{s} and \SI{59.1}{s}.}
\label{fig:3}
\end{figure}

The Marangoni velocity $V_\mathrm{M}$ at the side of the drop and velocity one radius above the drop $V_\mathrm{R}$ are measured (at positions as indicated in Fig.\ref{fig:2}(c)) and plotted in Fig.\ref{fig:3}(a). Here $V_\mathrm{R}$ represents the combined effect of the buoyancy jet and the Marangoni flow. The velocity of the drop $V_\mathrm{drop}$ is also plotted as a reference. Though $V_\mathrm{M}$ is slowly decreasing, $V_\mathrm{buo}$ decreases at a much higher rate, so that $V_\mathrm{R}$ decreases {to} zero and then changes direction (vertical dashed line). Shortly after, $V_\mathrm{M}$ increases, entering the positive {feedback} regime. The Marangoni flow is proportional to the surface tension difference on the drop, $V_\mathrm{M} \propto \Delta \sigma = 2R ({\partial \sigma}/{\partial w_\text{e}})({\partial w_\text{e}}/{\partial y}) \approx{\partial \sigma}/{\partial w_\text{e}}\cdot(w_\mathrm{e,B}-w_\mathrm{e,A})$, where $w_\mathrm{e,A}$ and $w_\mathrm{e,B}$ are the ethanol fractions of the surrounding mixture at the apex and bottom of the droplet, respectively (Fig.\ref{fig:2}(d)). The drop jumps from regions of ethanol fraction less than \SI{30}{wt\%}, below which, according to Fig.\ref{fig:1}(e), the interfacial tension decreases linearly, so that $\partial \sigma/\partial w_\mathrm{e}$ is a negative constant, thus $V_\mathrm{M} \propto  (w_\mathrm{e,A}-w_\mathrm{e,B})$. Consider the initial accelerating period where ethanol does not reach the drop's bottom yet, so only $w_\text{e,A}$ changes. Then $\text{d}V_\mathrm{M}/\text{d}t \propto \text{d}w_\mathrm{e,A}/\text{d}t$. In the region where the droplet jumps, the concentration of ethanol above the drop can be approximated as linear, $w_\mathrm{e} \propto y$, so that $\text{d}w_\mathrm{e,A}/\text{d}t \propto \text{d}y/\text{d}t$. The flow field at this moment is induced by the dominating Marangoni flow, so $\text{d}y/\text{d}t \propto V_\text{M}$, thus $\text{d}V_\mathrm{M}/\text{d}t \propto V_\mathrm{M}$. This gives rise to an exponential growth of the Marangoni flow:
\begin{equation}
V_\mathrm{M} \sim e^{t/\tau}
\end{equation}

A zoomed in logarithmic plot of $V_\mathrm{M}$ is shown in Fig.\ref{fig:3}(b). Indeed, $V_\mathrm{M}$ is confirmed to increase exponentially shortly after it starts to increase. The calculated {time constant $\tau$ of the growth} is \SI{0.073}{s}, fitting well with the measured value \SI{0.081}{s} (see Supplemental Material for coefficients). Note that the Marangoni flow remarkably increases by more than two orders of magnitude within \SI{1}{\second}, accounting for the sudden shooting up. 

$V_\mathrm{M}$ keeps increasing until the drop reaches a higher position where the ethanol fraction $w_e > \SI{40}{wt\%}$ because $\partial \sigma/\partial w_\mathrm{e}$ decreases sharply in this region, until it almost vanishes at $w_e \approx \SI{60}{wt\%}$ (Fig.\ref{fig:1}(e)), corresponding to $h\approx \SI{6}{mm}$, which forms the ``ceiling'' for the jumping because the Marangoni driving force ceases. $V_\mathrm{M}$ also decreases because when $V_\mathrm{M}$ reaches its highest value of $\sim \SI{5}{mm/s}$; the drop's P\'eclet number is then on the order of 1000. This strong advection tends to homogenize the surrounding liquid \cite{michelin2014phoretic}, leading to an additional decrease of ${\partial w_\text{e}}/{\partial y}$. This explains the formation of the almost uniform drifted liquid layer (Fig.\ref{fig:2}(a)). For a newly released drop, Marangoni flow during the injection process is responsible for the formation of the uniform layer. Note that when the droplet is sinking, the drifted layer around the drop also decreases in its ethanol concentration (Fig.\ref{fig:2}) because of advection and diffusion. This means that the interfacial surface tension of the droplet is building up, and the deeper it sinks, the more interfacial energy it builds up, so that the droplet has more energy to jump a larger distance. This accumulated interfacial energy is later transformed to momentum by Marangoni flow, in a kind of ``avalanche'' process.

\begin{figure}[t!] 
\centering
\includegraphics[width=0.48\textwidth]{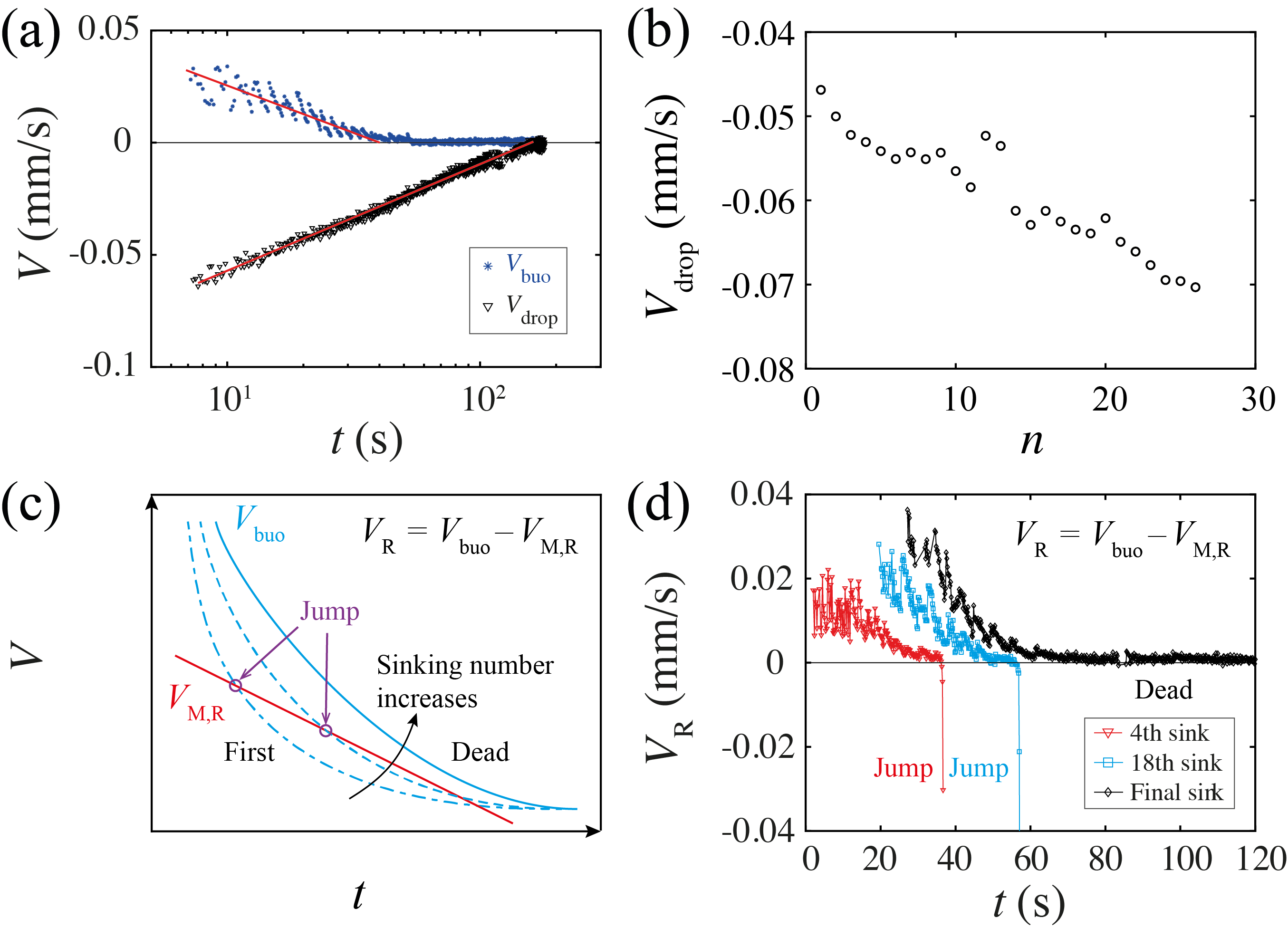}
\caption{(a) $V_\text{drop}$ and $V_\text{buo}$ (measured 1.5 radius above the drop) of a sinking droplet after the addition of surfacetant to the bulk liquid. (b) $V_\text{drop}$ at $h=\SI{2.5}{mm}$ as a function of the number of the subsequent sinking event. (c) Sketch of the balance between buoyancy jet and Marangoni flow. For later sinking events $n >1$, $V_\text{buo}$ changes from the blue dash-dotted line, to the dashed line and finally to the solid line. Because the last line shown does not intersect with $V_\text{M,R}$, so $V_\text{R}=V_\text{buo}-V_\text{M,R}$ does not change direction and the drop does not jump. (d) Measured $V_\text{R}(t)$ for different sinking cycles. The starting point (highest point) of each cycle is at $t=0$, but the initial stages of the cycles couldn't be measured because the camera could not record the entire cycle at the necessary frame rate.}
\label{fig:4}
\end{figure}

The direction of the flow above the droplet – which is determined by the relative strength between buoyancy jet and Marangoni flow – determines whether it jumps or not. To gain more insight into the buoyancy jet here, PIV measurements are performed {after the addtion of} \SI{0.7}{mM} SDS to suppress the Marangoni flow. It is found that both the sinking velocity $V_\text{drop}$ and the buoyancy jet velocity $V_\text{buo}$ (measured 1.5 radius above the droplet) decrease exponentially, as shown in Fig.\ref{fig:4}(a), with the former one similar to that of a sinking particle \cite{zvirin1975settling}. Meanwhile, $V_\text{M}$ decreases at a much slower rate than $V_\text{buo}$ (Fig.\ref{fig:2}(a)), so it is most likely that Marangoni flow will dominate when $V_\text{drop}$ is small enough. However, the Marangoni flow is getting weaker by each jump (see Supplemental Material), most likely due to mixing of the surrounding liquid \cite{blanchette2012drops}, which is enhanced by the jumping itself. Therefore the Marangoni-flow-induced lifting force gets smaller at each jump, so that the sinking velocity increases, as shown in Fig.\ref{fig:4}(b), where the sinking velocity at $h=\SI{2.5}{mm}$ is plotted against the subsequent number of sinks. Consequently, $V_\text{buo}$ increases, and therefore it gets progressively harder for the Marangoni flow to overcome buoyancy. Fig.\ref{fig:4}(c) shows a sketch of the \textit{relative} strength between $V_\text{buo}$ and $V_\text{M,R}$ (Marangoni flow induced flow 1.5 radius above the drop). For the first sink, $V_\text{buo}$ decreases quite fast, and the flow reversal happens when $V_\text{M,R}$ becomes larger than $V_\text{buo}$, so that the droplet jumps. As the number of sinking events increases, $V_\text{buo}$ increases and $V_\text{M,R}$ decreases. Therefore the droplet can sink longer (and also deeper) before it jumps. Finally at some point, $V_\text{M,R}$ is so weak that $V_\text{buo}$ is always dominant: The jumping stops and the droplet falls ``dead''. We confirm this picture by measuring $V_\text{R}=V_\text{buo}-V_\text{M,R}$ for a normal sinking droplet without surfactant (Fig.\ref{fig:4}(d)). As expected, $V_\text{R}$ decreases from a higher value and changes direction at later times as increasing sinking number, until finally no flow reversal is observed, and the droplet falls dead.

In conclusion, an oil droplet of trans-Anethole released in a vertically stratified ethanol/water mixture is found to bounce repeatedly with ever increasing jumping distance, until finally it falls dead all of a sudden. Marangoni flow and gravity are responsible for this phenomenon: Interfacial energy builds up when the droplet sinks, and a flow reversal above the sinking droplet triggers an exponential growth of the Marangoni flow, leading to the sudden jump. The consequent strong advection decreases the Marangoni stress, enabling the droplet to sink again and then continue the bouncing cycle. The ever decreasing Marangoni flow by each  jump is responsible for the droplet's increasing jumping height as well as its sudden death. 

The present system can be easily generalized to other liquids, as long as one has a vertically stratified liquid which can generate strong enough Marangoni stress on the droplet. This is supported by our observation of a silicon oil drop in the same stratified fluid which exhibits a similar bouncing behavior. In addition to potential applications for oil recovery and drug delivery, this new type of bouncing may also pave a new way for droplet manipulation and micromixing. 

Valuable discussions with Chao Sun, Xiaojue Zhu, Guillaume Lajoinie, Huanshu Tan, Yaxing Li and Luoqin Liu are greatly appreciated. G. Lajoinie also provided valuable technical support. We acknowledge support from the Netherlands Center for Multiscale Catalytic Energy Conversion (MCEC), an NWO Gravitation programme funded by the Ministry of Education, Culture and Science of the government of the Netherlands, and ERC-Advanced Grant under project number 30012101. X. H. Z. also acknowledges support from Discovery Project and Canada Research Chair program from Natural Sciences and Engineering Research Council of Canada.

\end{document}